\documentclass[a4paper,11pt]{article}

\usepackage{jheppub} 
\usepackage{tikz}

\usepackage{amsmath}
\usepackage{amssymb}
\usepackage{braket}
\usepackage{comment}
\usepackage[many]{tcolorbox}
\usepackage{enumerate}
\usepackage[multiple]{footmisc}

\newtcolorbox{empheqboxed}{colback=gray!30, 
 colframe=white,
 width=\textwidth,
 sharpish corners,
 top=-2mm, 
 bottom=0mm
}

\newcommand{\TT}{T\overline{T}}

\newcommand{\ul}{\underline}

\def\AdS{{\rm AdS}}

\newcommand{\Z}{{\mathbb Z}}

\newcommand{\ap}{\alpha'}

\newcommand{\bth}{\beta_{\text{th}}}
\newcommand{\bcd}{\beta_{\text{cd}}}
\newcommand{\ALD}{\rm ALD}

\newcommand{\C}[1]{$(\ref{#1})$}

\title{The On-shell Gravity Action and Linear Dilaton Holography}

\author[a]{Andrea Dei,}
\author[b]{Kiarash Naderi,}
\author[a]{Savdeep Sethi}
\affiliation[a]{\vskip 0.01cm
Leinweber Institute for Theoretical Physics \& Enrico Fermi Institute \& Kadanoff Center for Theoretical Physics, University of Chicago, Chicago, IL 60637, USA}
\affiliation[b]{Institut f\"ur Theoretische Physik, ETH Z\"urich,\\ 
Wolfgang-Pauli-Strasse 27, 8093 Z\"urich, Switzerland}
\emailAdd{adei@uchicago.edu}
\emailAdd{knaderi@phys.ethz.ch}
\emailAdd{sethi@uchicago.edu}

\abstract{Computing the Euclidean spacetime action on-shell provides a useful way of both testing holographic proposals and determining the string theory sphere partition function. We consider families of three-dimensional linear dilaton spacetimes for which there are holographic proposals that share features of a $\TT$-deformed CFT.  We extend the holographic renormalization program beyond AdS to this class of geometries by identifying the boundary terms needed for a well-defined variational principle and a finite on-shell action. We show that the spacetime energy or mass determined from the on-shell action matches the $\TT$-deformed two-dimensional CFT energy. This provides more evidence for the role of the $\TT$ deformation in this holographic correspondence.}

\begin{document}

\maketitle

\newpage

\section{Introduction and summary} \label{sec:introduction}

\subsubsection*{\ul{\it Motivation}}

Any theory of quantum gravity that admits black holes is expected to be holographic. One way to learn about the holographic definition is by studying the quantum gravity action evaluated on interesting solutions like Euclidean black holes. The value of the Euclidean action is then related to the mass of the spacetime and to the Bekenstein-Hawking entropy \cite{Gibbons:1976ue, Hawking:1982dh}. From the Euclidean action, we can therefore learn a great deal about the states of a dual holographic description. 

Our aim in this work is to compute the Euclidean quantum gravity action for a class of three-dimensional spacetimes which are neither asymptotically flat nor asymptotically $\AdS$. Rather, we are interested in spacetimes which are asymptotically linear dilaton (ALD). Very loosely, this means the scalar field that we choose to call the dilaton, denoted $\Phi$, has a linear dependence on a coordinate near infinity. This can be a spacelike, timelike or null coordinate. In the examples we will study here, the backgrounds are $\AdS_3$ spacetimes in the deep bulk. This will provide an important renormalization condition to help determine the needed boundary terms. The dilaton is linear in a spacelike radial coordinate near spatial infinity. The corresponding gradient energy for this scalar field changes the structure of the boundary, breaking the conformal symmetry of an $\AdS_3$ spacetime.

The reason for our interest in this class of spacetimes is that they arise naturally in string theory. We will therefore use the string effective action to define the bulk theory. This amounts to adding some matter fields in addition to the metric. More importantly, we have recently learned more about the holographic definition of quantum gravity on spacetimes with \ALD$_3$ asymptotics. The holographic dual has many similarities to a $\TT$-deformed CFT~\cite{Giveon:2017nie, Giveon:2017myj, Asrat:2017tzd, Hashimoto:2019wct, Hashimoto:2019hqo, Apolo:2019zai, Chakraborty:2020yka, Chang:2023kkq}. While the precise role of the $\TT$ deformation is still under discussion for generic string tension, for the special case of the tensionless string the holographic boundary theory is a single trace $\TT$-deformed symmetric product CFT~\cite{Dei:2024sct, Dei:2025ilx}. 

Computing the Euclidean action typically involves a regulation procedure to get a finite result. In $\AdS$ spacetimes one regulation procedure, called holographic renormalization, is the gravity version of the renormalization procedure of quantum field theory; for a review, see~\cite{Skenderis:2002wp}. In more recent work, the procedure of holographic renormalization has been extended beyond asymptotically AdS spacetimes \cite{Kraus:1999di, Cai:1999xg, Mann:2005yr, Mann:2006bd, Astefanesei:2006zd, Marolf:2007ys, LeWitt:2008zx, Mann:2008ay, Kanitscheider:2008kd, Wiseman:2008qa, Mann:2009id}. The primary issue is determining which boundary terms to add to a bulk quantum gravity action describing spacetimes with some fixed asymptotic behavior. Given a choice of asymptotic behavior, the choice of boundary terms is determined, in part, by requiring a well-defined variational principle; namely that the Euclidean action is stationary on spacetimes with those desired asymptotics~\cite{Regge:1974zd, Banados:1992wn, Papadimitriou:2005ii, Mann:2005yr, Marolf:2007ys, LeWitt:2008zx, Kanitscheider:2008kd, Mann:2009id}. As we will see, satisfying this requirement will still leave some ambiguity in the choice of boundary couplings, which must be resolved using additional physical inputs.  

\subsubsection*{\ul{\it Connection to string theory}}

A second motivation for this spacetime computation is that it provides a way to evaluate the string theory sphere partition function for these specific backgrounds. 
In particular, the supergravity solutions we consider are bound states of NS5-branes and fundamental strings studied in the decoupling limit where the asymptotic string coupling $g_s \rightarrow 0$. The key point of this partial decoupling limit is that the string scale $\alpha'$ is held fixed. The resulting backgrounds take the form $\mathcal M_3 \times \text S^3 \times \mathbb T^4$, where $\mathcal M_3$ interpolates between an AdS$_3$ solution and linear dilaton asymptotics. Taking the low-energy limit, or equivalently $\alpha' \rightarrow 0$ results in an AdS$_3$ solution. We discuss these solutions in detail in Section~\ref{sec:ald}. 

To evaluate the partition function, we restrict to cases where the string coupling is weak everywhere so the string partition function can be computed in a genus expansion. The first term in this expansion, $Z_{\text{sphere}}$, is notoriously difficult to compute because one needs to divide by the volume of the global conformal group $SL(2,\mathbb{C})$. In practice, this means regulating the ratio of the infinite volume of the target space geometry divided by the infinite volume of  $SL(2,\mathbb{C})$. The sphere partition function for particular cases like global AdS$_3$ with pure NS-NS flux and certain two-dimensional gravity models have proven amenable to worldsheet CFT techniques~\cite{Anninos:2021ene, Mahajan:2021nsd, Eberhardt:2023lwd}. It would be very interesting to extend those kinds of approaches to ALD backgrounds, although some ingredients which appear necessary are still missing. 

For weakly-curved non-compact target spaces, the more common approach is to determine the sphere partition function by evaluating the target space Euclidean tree-level effective action,
\begin{align}
    S_{\text{on-shell}} =- \frac{1}{2 \kappa_{10}^2}\int \text d^{10}x \, e^{-2\Phi} \sqrt{g} \, R + \ldots \, , 
\end{align}
where the omitted terms include other fields as well as higher derivative interactions. The on-shell value of the tree-level effective action should reproduce the sphere partition function via,
\begin{equation} \label{eq:intro-sphere-pt-fn}
Z_{\text{sphere}}=e^{-S_\text{on-shell}}\ . 
\end{equation}
Here by $S_\text{on-shell}$ we mean the tree-level bulk effective action together with all boundary terms needed for at least a well-posed variational problem. Truncating to the two derivative spacetime effective action gives $Z_{\text{sphere}}$ to leading order in a weak curvature or $\alpha'$ expansion. For a recent computation using this approach, see~\cite{Ahmadain:2024hgd}. The advantage of the spacetime approach is that it can be studied for a large class of string backgrounds, whether they admit a tractable worldsheet description or not. The disadvantage is the restriction to weak curvature. 

\subsubsection*{\ul{\it Holography beyond AdS/CFT}}

The reason to focus on ALD spacetimes is that they offer a nice intermediary between asymptotically AdS and asymptotically flat geometries. Unlike $\AdS$, asymptotically linear dilaton spacetimes have a null boundary. Correspondingly, the physics of large black holes will be quite different from AdS spacetime, where large black holes can exist in stable equilibrium with their own Hawking quanta. In this case, some of those quanta will escape to null infinity. For the ALD$_3$ case, the high-energy density of states dominated by black holes exhibits Hagedorn rather than Cardy-like growth. This is the basic reason the holographic dual cannot be a local quantum field theory but something more exotic like a $\TT$-deformed field theory. 

For weakly-coupled string theories with a holographic dual, the sphere partition function is expected to compute the leading energy of the state dual to the chosen background. For example, the on-shell action for $\text{AdS}_5 \times \text{S}^5$ regulated using holographic renormalization reproduces the Casimir energy of $\mathcal N=4$ Super Yang Mills \cite{Balasubramanian:1999re}. Similarly from the thermal AdS$_3$ result, one can correctly extract the $E_0 =-\frac{c}{12 L}$ vacuum energy of the boundary CFT on a cylinder of size $L$ \cite{Brown:1986nw,Henningson:1998gx}. In the same spirit, our goal is to regulate the on-shell action for the ALD$_3$ solutions reviewed in Section~\ref{sec:ald} to compute the sphere partition function to leading order in $\alpha'$. From this result, we can deduce the energy of the associated boundary state. 

These geometries interpolate between AdS$_3$ and asymptotically linear dilaton. The $\AdS_3$ has a CFT$_2$ dual description. The boundary theory is believed to share features with the $\TT$ deformation of this CFT$_2$ dual. One might then suspect that the ALD$_3$ spacetime mass is the $\TT$-deformed CFT$_2$ energy. In Section~\ref{sec:on-shell} we confirm that this is indeed the case; see eq.~\eqref{eq:e-lambda-ttbar}.

\subsubsection*{\ul{\it Outline and future directions}}

The paper is organized as follows. In Section~\ref{sec:dimensional-reduction}, we dimensionally reduce the Type II supergravity action and obtain a three-dimensional effective action; see eqs.~\eqref{S3-final-Einstein} and \eqref{V_E}.  We proceed in Section~\ref{sec:ald} by reviewing the ALD$_3$ solutions with metrics given in~\eqref{eq:ald-sols} and studying their regularity. Identifying the needed boundary terms is the subject of Section~\ref{sec:logic}. The final result is given in eq.~\eqref{eq:ald-bdy-terms-final}. 

Section~\ref{sec:on-shell} is the core of the paper: we compute the on-shell action for different values of the parameters entering the partially decoupled solutions. We determine the mass of the spacetime from this computation. As one should expect \cite{Iyer:1995kg}, we verify that our result obtained from this Euclidean computation agrees with the mass of the spacetime previously computed using the covariant phase space formalism~\cite{Chang:2023kkq}. 
We discuss how our findings are in agreement with the thermodynamic expectations and reproduce the Bekenstein-Hawking entropy for the black hole solutions. Finally, we discuss the holographic interpretation and the relation of the sphere partition function to the $\TT$ deformation. 

Our results build on interesting past work studying holographic renormalization in ALD spacetimes with spacetime dimension $D>3$~\cite{Marolf:2007ys, Mann:2009id}. 
There has also been past work exploring possible boundary terms for the bulk gravity action for models related to the standard double trace~\cite{Tian:2023fgf, FarajiAstaneh:2024fig}, single-trace $\TT$-deformed CFT~\cite{Chakraborty:2020swe, Chakraborty:2022dgm} and $\TT$-like flows in two dimensions~\cite{Grumiller:2020fbb}. 
What we are leveraging in this work is the existence of a limit where we must recover known AdS results. For general ALD backgrounds, such a limit need not exist. 
With our improved understanding of how to compute the sphere partition function for ALD geometries with AdS interiors, one can apply similar techniques to a number of problems from spinning black holes in these ALD$_3$ geometries, where the covariant phase space approach encounters integrability issues~\cite{Chang:2023kkq}, to string solutions in other dimensions with similar qualitative features like~\cite{Martinec:2017ztd, Ferko:2025elh}. 

\section{Dimensional reduction} 
\label{sec:dimensional-reduction}

The goal of this section is to dimensionally reduce the ten-dimensional pure NS-NS Type II supegravity action on $\text S^3 \times \mathbb T^4$ to obtain an effective action in three dimensions. The final expression is given in eqs.~\eqref{S3-final-string} and \eqref{VS-sol} for  string frame and in eqs.~\eqref{S3-final-Einstein} and \eqref{V_E} for Einstein frame. The reader not interested in the derivation of these results can skip directly to Section~\ref{sec:ald}. 

\subsection{Three-dimensional effective action in string frame}

We consider geometries of the form $\mathcal M_3 \times \text S^3 \times \mathbb T^4$, where $\mathcal M_3$ is three-dimensional and Euclidean. For the metric, NS-NS three-form flux $H_3$ and dilaton $\Phi$ we assume the following ansatz,\footnote{The somewhat unfamiliar factor of $i$ in the expression for $H_3$ is a consequence of the Wick rotation that maps Lorentzian solutions to Euclidean solutions.} 
\begin{subequations}
\label{dim-red-ansatz-initial}
\begin{align}
     & \text ds^2  = \text ds_{\mathcal M_3}^2 + n_5 \, \alpha ' \,  \text ds_{\text S^3}^2 + \text ds^2_{\mathbb T^4} \ , \\
    & H_3 = 2 i f(\Phi) \, \text{vol}(\mathcal M_3) + 2 n_5  \alpha'  \text{vol}(\text S^3)  \ , \qquad 
    \Phi = \Phi(r) \,,
\end{align}
\end{subequations}
where $f(\Phi)$ is a function of the dilaton $\Phi$ and $r$ is one of the coordinates parameterizing $\mathcal M_3$. We assume no R-R flux is present and we restrict to solutions where the metric $\text ds^2_{\mathcal M_3}$ only depends on $r$, which will be a radial coordinate. 
In \eqref{dim-red-ansatz-initial}, $\text ds_{\text S^3}^2$ denotes the unit three-sphere metric  and $\text{vol}(\text S^3)$ is the associated volume form with $\int_{\text S^3} \text{vol}(\text S^3) = 2 \pi^2$. 
The radius of the three-sphere is fixed with $R_S^2 = n_5\alpha' $ by requiring flux quantization, 
\begin{equation} \label{eq:h3-flux}
    \frac{1}{4\pi^2\alpha'} \int_{\text S^3} H_3 = n_5\ ,  \qquad  n_5\,\in\Z\ . 
\end{equation}
Finally, in our conventions the volume of $\mathbb{T}^4$ is parametrized as $\text{vol}(\mathbb T^4) = (2\pi)^4 \, \ap^2 \, v_4$.

In order to derive the three-dimensional effective action for the class of backgrounds in eq.~\eqref{dim-red-ansatz-initial}, we begin by considering the string-frame ten-dimensional bosonic action for Type II supergravity in absence of R-R fluxes,
\begin{equation}
    S_{10} = -\frac{1}{2\kappa_{10}^2}\int \text d^{10}x \, \sqrt{g} \, e^{-2 \Phi}\Bigl(R + 4 \, \partial_\mu \Phi \partial^\mu \Phi - \frac{1}{12}H_{\mu \nu \rho}H^{\mu \nu \rho} \Bigr) \ , 
    \label{S10}
\end{equation}
where
\begin{equation}
   2 \kappa^2_{10} =   (2\pi)^7 \alpha'^{4} \ .  
\end{equation}
Varying the supergravity action \eqref{S10} one obtains the string-frame equations of motion  
\begin{subequations}
\label{10dsugra}
\begin{align}
    & \nabla^\mu( e^{-2 \Phi}  H_{\mu \nu \rho})=0 \ , \label{10dsugra-1}\\
    & R_{\mu \nu} + 2 \nabla_\mu \nabla_\nu \Phi - \frac{1}{4} H_{\mu \rho \sigma} H_\nu{}^{\rho \sigma} = 0 \ , \label{10dsugra-2} \\
    & R + 4 \nabla_\mu \nabla^\mu \Phi - 4 \nabla_\mu \Phi \nabla^\mu \Phi -\frac{1}{12}H_{\mu \nu \rho}H^{\mu \nu \rho} = 0 \ .   \label{10dsugra-3}
\end{align}
\end{subequations}
Specializing to the class of geometries \eqref{dim-red-ansatz-initial} reduces eq.~\eqref{10dsugra-1} to $\nabla^r(e^{-2\Phi} f(\Phi)) = 0$ and we therefore find that the only possibility is\footnote{From now on, for simplicity, we will assume $n_1, n_5 > 0$.}
\begin{equation}
    f(\Phi) = \frac{n_1}{\ap^{1/2} \, n_5^{3/2} \, v_4} \, e^{2 \Phi} \ ,  
    \label{f-sol}
\end{equation}
where we fixed the integration constant by requiring $H_7 = e^{-2\Phi} \, *_{10} \,  H_3 $ be quantized, 
\begin{equation}
  \frac{1}{ (2\pi)^6 (\alpha')^3}  \int_{\text S^3 \times \mathbb T^4} H_7 = n_1\ , \qquad  n_1\in\Z\ .  
\label{H7-quantization}
\end{equation}
We can thus rewrite eqs.~\eqref{dim-red-ansatz-initial} as 
\begin{subequations}
\label{dim-red-ansatz}
\begin{align}
    & \text ds^2  = \text ds_{\mathcal M_3}^2 + n_5 \alpha ' \,  \text ds_{\text S^3}^2 + \text ds^2_{\mathbb T^4} \ , \\
    & H_3 =  \frac{2i n_1}{\ap^{1/2} \, n_5^{3/2} \, v_4} \, e^{2 \Phi} \, \text{vol}(\mathcal M_3) + 2n_5  \alpha' \, \text{vol}(\text S^3)  \ , \qquad
    \Phi = \Phi(r) \ .  
\end{align}
\end{subequations}
Specializing eqs.~\eqref{10dsugra-2} and \eqref{10dsugra-3} to the ansatz \eqref{dim-red-ansatz}, we find
\begin{subequations}
\label{EoM-reduced}
\begin{align}
    & \hat R_{\mu \nu} + 2 \hat \nabla_\mu \hat \nabla_\nu \Phi + \frac{2 \, n_1^2}{\ap \, v_4^2 \, n_5^3} \, \hat g_{\mu \nu} \, e^{4 \Phi} = 0 \ ,   \label{EoM-reduced-1} \\
    & \hat R - 4 \hat \nabla_\mu \Phi \hat \nabla^\mu \Phi + 4 \hat \nabla_\mu \hat \nabla^\mu \Phi + \frac{4}{\alpha' \, n_5} + \frac{2 \, n_1^2}{\ap \, v_4^2 \, n_5^3} \, e^{4 \Phi} = 0 \ , \label{EoM-reduced-2}
\end{align}
\end{subequations}
respectively. In eqs.~\eqref{EoM-reduced}, in order to distinguish them from their ten-dimensional analogues, we introduced hatted symbols to denote the three-dimensional metric $\hat g_{\mu \nu}$ on $\mathcal M_3$, the associated Ricci tensor $\hat R_{\mu \nu}$, Ricci scalar $\hat R$ and so on. 

One can check that the three-dimensional equations of motion \eqref{EoM-reduced} can be derived from the string-frame effective action
\begin{equation}
S_{\text{bulk}}^{(\text{str})} = - \frac{1}{2 \kappa_3^2} \int \text d^3x \, \sqrt{\hat g} \, e^{-2 \Phi}\left(\hat R + 4 \, \partial_\mu \Phi \partial^\mu \Phi - V^{(\text{str})}(\Phi) \right) \ , 
\label{S3-final-string}    
\end{equation}
where the potential reads
\begin{equation}
    V^{(\text{str})}(\Phi) = \frac{1}{\ap} \left( \frac{2 n_1^2}{  v_4^2  n_5^3} \, e^{4 \Phi } - \frac{4}{  n_5} \right) \ ,  
    \label{VS-sol}
\end{equation}
and the normalization factor $2 \kappa^2_3$ can be expressed in terms of fundamental constants as follows,
\begin{equation} \label{eq:kappa3-value}
    \frac{1}{2\kappa_3^2} =  \frac{2 \pi^2 R_S^3 \, \text{vol}(\mathbb T^4)}{2\kappa_{10}^2} = \frac{n_5^{3/2} \, v_4}{4 \pi \alpha'^{1/2}} \ .  
\end{equation}

\subsection{Three-dimensional effective action in Einstein frame}

In our subsequent discussion, it will often be useful to work in Einstein frame, which is related to the string frame via 
\begin{equation} \label{stringtoeinstein}
    \hat g_{\mu \nu} = e^{-4 \Phi} \hat g^{(\text{str})}_{\mu \nu} \ .  
\end{equation} 
The relation \C{stringtoeinstein} differs from the usual transformation between frames in a subtle but important way. Usually the dilaton tends to a finite value, $\langle \Phi \rangle$, at spatial infinity and one only uses $\Phi - \langle \Phi \rangle$ to change frames with the aim of keeping the asymptotic gravitational interaction strength fixed. In this work, however, we want to allow $\Phi$ to go to a boundary of moduli space at spatial infinity. This shuts off the gravitational interaction at spatial infinity. For these special backgrounds, it is more useful to change frames using the full dilaton appearing in \C{stringtoeinstein}, including the zero mode, when relating string and Einstein frames.  

We obtain the three-dimensional effective action in Einstein frame\footnote{In transforming between string and Einstein frame, we are neglecting boundary terms that do not affect the equations of motion, which are the focus of this discussion. In Section~\ref{sec:ald-on-shell-comptutation}, we are going to analyze which boundary terms are necessary for the variational principle to be well-posed and for the on-shell action to be finite.} 
\begin{equation}
      S_{\text{bulk}} = - \frac{1}{2 \kappa_3^2} \int \text d^3x \,  \sqrt{\hat g}\bigl(\hat R - 4 \partial_\mu \Phi \partial^\mu \Phi - V(\Phi) \bigr) \ , 
      \label{S3-final-Einstein}
\end{equation}
where
\begin{equation}
    V(\Phi) =  \frac{1}{\ap} \left( \frac{2 n_1^2}{ v_4^2 n_5^3} \, e^{8 \Phi} - \frac{4}{ \, n_5} e^{4 \Phi} \right) \ .  
    \label{V_E}
\end{equation}
The associated Einstein frame equations of motion read:
\begin{subequations}
\begin{align}
    & \hat R_{\mu \nu} - 4 \, \partial_\mu \Phi \partial_\nu \Phi  - \hat g_{\mu \nu} V(\Phi)=0 \ , \\ 
    & \hat \nabla_\mu \hat \nabla^\mu \Phi -\frac{1}{8}V'(\Phi) = 0 \ .  
\end{align}
\label{EoM-3d-E}
\end{subequations}

\section{Partially decoupled solutions } 
\label{sec:ald}

As we mentioned in the Introduction, one way to construct geometries interpolating from AdS$_3$ in the deep interior to linear dilaton asymptotics is to partially decouple a bound state of NS5-branes and  fundamental strings~\cite{Seiberg:1997zk,Aharony:1998ub,Giveon:1999zm,Tseytlin:1996as,Chang:2023kkq}.  
In this section, we consider such a solution of the form $\mathcal{M}_3 \times \text{S}^3 \times \mathbb{T}^4$, see eq.~\eqref{dim-red-ansatz}. Specifically we consider a Euclidean string frame metric on $\mathcal{M}_3$ obtained by continuing a non-spinning black hole using a parametrization based on~\cite{Chang:2023kkq},
\begin{subequations} 
\label{eq:ald-sols}
    \begin{equation} 
    \label{eq:main-metric}
	\text ds^2_{\mathcal M_3} = \frac{r^2-8 M \ap n_5}{r^2 + R_\phi^2} \, \frac{R_\phi^2}{\ap n_5} \, \text dt^2 + \frac{r^2 }{r^2 + R_\phi^2} \, R_\phi^2 \, \text  d\phi^2 + \frac{\ap n_5}{r^2-8 M \ap n_5} \, \text dr^2 \,.
    \end{equation}
The associated dilaton reads
    \begin{equation} 
    \label{eq:main-dilaton}
	e^{2\Phi} = \frac{v_4 \, n_5 \, R_\phi^2 \sqrt{1+\frac{8 M \ap n_5}{R_\phi^2}}}{n_1 (r^2 + R_\phi^2)} \ .  
    \end{equation}
\end{subequations}
After passing to Einstein frame, it is easy to check that eqs.~\eqref{eq:ald-sols} solve the equations of motion~\eqref{EoM-3d-E} with the potential $V(\Phi)$ given in eq.~\eqref{V_E}. 

Let us explain the various coordinates and parameters entering eq.~\eqref{eq:ald-sols}. The coordinates $(t,r)$ have length dimension one, while $\phi$ is dimensionless and compact: $\phi \sim \phi+2\pi$. The integers $n_5$ and $n_1$ determine the number of NS5-branes and F1 strings through the flux quantization conditions~\eqref{eq:h3-flux} and \eqref{H7-quantization}. The parameter $R_\phi$ has units of length and controls the scale at which the crossover between AdS$_3$ and linear dilaton geometry occurs. 

In fact, we show in the next subsection that the limit $R_\phi \to \infty$ recovers asymptotically AdS$_3$ solutions. In this limit, $M>0$ parametrizes the black hole mass in units of the three-dimensional Newton constant $G_3$. The parameter $v_4$ is related to the volume of $\mathbb{T}^4$; see the comments below eq.~\eqref{eq:h3-flux}. 

\paragraph{Normal coordinates and asymptotic fall-off.}
In the following sections, it will be useful to work with a canonical normal coordinate $\eta$ instead of $r$. We therefore rewrite the three-dimensional Einstein-frame metric \eqref{eq:main-metric} as follows,
\begin{equation}
    \text ds^2 = \hat g_{\mu \nu} \, \text dx^\mu \text dx^\nu = \text d\eta^2 + h_{ij}(\eta) \, \text dx^i \text dx^j \ .
    \label{eq:ald-normal}
\end{equation}
One can then verify that the asymptotic fall-off of the metric and dilaton for large $\eta$ reads \begin{align}
    e^{2\Phi(\eta)} &= \frac{\sqrt{\ap n_5}}{2} \, \eta^{-1} + \mathcal{O}(\eta^{-2}) \ , \label{Phi-falloff} \\
    h_{ij}(\eta) &=  h_{ij}^{(0)} \, \eta^2 + \mathcal{O}(\eta) \ , \label{h-falloff}
\end{align}
where $h^{(0)}_{ij}$ does not depend on $\eta$. 

\subsection[Asymptotically AdS$_{3}$ solutions]{Asymptotically AdS$_{\boldsymbol 3}$ solutions}
\label{sec:AAdS}

The limit $R_\phi \to \infty$ implements the further decoupling of fundamental strings and leads to asymptotically AdS$_3$ geometries. In this limit the dilaton is constant,
\begin{equation}
    e^{2\Phi} = \frac{n_5}{n_1} \, v_4 \ ,   \label{eq:ads3-main-dilaton}
\end{equation}
and the string-frame effective action \eqref{S3-final-string} reduces to 
\begin{equation}
    S_{\text{AdS}_3} = - \frac{1}{16 \pi G_3} \int \text d^3x \, \sqrt{\hat g}\left( \hat R + \frac{2}{R_{\text{AdS}}^2} \right) \ , 
    \label{AdS3-bulk-action}
\end{equation}
where we introduced the three-dimensional Newton's constant $G_3$,
\begin{equation} \label{eq:g3-newton}
\frac{1}{16 \pi G_3} = \frac{e^{-2 \Phi}}{2 \kappa^2_3} = \frac{n_1}{4 \pi} \sqrt{\frac{n_5}{\ap}}   \ , 
\end{equation}
and 
\begin{equation} \label{eq:rads3}
    R_{\text{AdS}} = \sqrt{\alpha' n_5}
\end{equation}
denotes the AdS$_3$ radius. Taking the limit $R_\phi \to\infty$ sends the geometry \eqref{eq:ald-sols} to
\begin{align} 
	\text ds^2_{\text{AdS}_3}  &= \Bigl(\frac{r^2}{R_{\text{AdS}}^2} - 8M \Bigr) \text dt^2 + r^2 \, \text d\phi^2 + \Bigl(\frac{r^2}{R_{\text{AdS}}^2} - 8M \Bigr)^{-1} \text dr^2 \ .  
    \label{eq:ads3-main-metric}
\end{align}
This solution describes different locally AdS$_3$ geometries depending on the mass parameter $8M \geq -1$, which we are now going to briefly review.\footnote{If $8M<-1$, the conical excess of the $\phi$ coordinate requires the presence of objects with negative tension. Therefore, one restricts to $8M \geq -1$.} 

\paragraph{$\boldsymbol{8M=-1}$: Global AdS$_{\boldsymbol 3}$ and thermal AdS$_{\boldsymbol 3}$.} For $M= -\frac{1}{8}$, the geometry \eqref{eq:ads3-main-metric} reduces to global AdS$_3$ with metric 
\begin{equation}
    \text ds_{\mathcal M_3}^2 = \Bigl(\frac{r^2}{R_{\text{AdS}}^2} +1 \Bigr) \text dt^2 + r^2 \text d\phi^2 + \Bigl(\frac{r^2}{R_{\text{AdS}}^2} +1 \Bigr)^{-1} \text dr^2 \ .  
    \label{global-AdS3}
\end{equation}
The Euclidean time coordinate $t$ is non-compact and the asymptotic boundary has the topology of a cylinder. The solution is regular everywhere and $r \geq 0$. 

Periodically identifying  the Euclidean time coordinate of global AdS$_3$ gives thermal AdS$_3$. The metric is still \C{global-AdS3} but the asymptotic boundary has the topology of a torus and Euclidean time is compactified,  
\begin{equation}
 t \sim t + \bth  \ .      
\end{equation}
Notice that the Euclidean time circle is not contractible, so there is no condition imposed on the periodicity of $\bth$. The inverse temperature $\bth$ is not related to the mass, which takes the same value $M= -\frac{1}{8}$ as global AdS$_3$. This solution is regular everywhere with $r\geq 0$. 

\paragraph{$\boldsymbol{-1 < 8M < 0}$: Conical defect geometries.} Conical defect geometries have masses $8M = - \frac{1}{n^2}$ with $n \in \mathbb Z$ \cite{Balasubramanian:2000rt, Maldacena:2000dr, Lunin:2002iz, Kraus:2006wn}. For $n = \pm 1$ this is global or thermal AdS$_3$. Rescaling coordinates in eq.~\eqref{eq:ads3-main-metric} using $t \mapsto n t$, $r \mapsto n^{-1} r$ and $\phi \mapsto n \phi$, we find the metric \eqref{global-AdS3}. As a consequence of the rescaling, the compact coordinate $\phi$ acquires a deficit angle, 
\begin{equation}
    \phi \sim \phi + \frac{2 \pi}{n} \ .  
\end{equation}
The Euclidean time coordinate can be compactified or not, corresponding to asymptotic boundaries with the topology of a torus or cylinder, respectively. Our following discussion is focused on geometries with a torus boundary, so we will compactify $t$ via $t \sim t + \bcd$. For $n \neq \pm 1$, conical defect geometries feature a time-like singularity at the origin. 

\paragraph{$\boldsymbol{M=0}$: Cusp geometry.} Let us then consider the case $M=0$. The metric \eqref{eq:ads3-main-metric} reduces to 
\begin{equation}
    \text ds^2_{\text{AdS}_3}  = \frac{r^2}{\ap n_5} \text dt^2 + r^2 \, \text d\phi^2 + \frac{\ap n_5}{r^2} \text dr^2 \ .  
\end{equation}
No horizon is present and the geometry is well-defined for $r>0$. At $r=0$ the Riemann tensor is diverging, although there are no diverging diffeomorphism invariants. Hence the terminology cusp singularity. One is free to choose whether the Euclidean time coordinate is compact or not, and as mentioned above, we will be interested in compact Euclidean time. 

\paragraph{$\boldsymbol{M > 0}$: BTZ black hole.} For $M>0$ we find the continuation of the BTZ black hole \cite{Banados:1992wn, Banados:1992gq}. The metric \eqref{eq:ads3-main-metric} can be rewritten as
\begin{equation}
    \text ds^2_{\text{AdS}_3}  = \frac{r^2-r_+^2}{R_{\text{AdS}}^2} \text dt^2 + r^2 \, \text d\phi^2 + \frac{R_{\text{AdS}}^2}{r^2-r_+^2} \text dr^2 \ , 
    \label{BTZ}
\end{equation}
where $r_+$ denotes the radial coordinate of the BTZ black hole event horizon, 
\begin{equation}
    r_+^2 = 8 M \ap n_5 \ .  
\end{equation}
In order for the geometry \eqref{BTZ} to be smooth and avoid conical singularities, the Euclidean time coordinate $t$ needs to be periodically identified, 
\begin{equation}
t \sim t + \beta_{\text{BTZ}} \ , \qquad \beta_{\text{BTZ}} = \frac{2 \pi R_{\text{AdS}}^2}{r_+} = 2\pi \sqrt{\frac{\ap n_5}{8 M}} \ .  
\label{BTZ-beta}
\end{equation}
Notice that both $t$ and $\phi$ are compact and the asymptotic boundary of the Euclidean BTZ black hole is a torus. The geometry caps off smoothly at the horizon $r=r_+$ and the Euclidean solution is only defined for $r \geq r_+$.

\paragraph{Normal coordinates.} As in the prior case, it is useful to consider a canonical normal coordinate \eqref{eq:ald-normal}. One can easily verify that the boundary metric $h_{ij}$ (for $R_{\text{AdS}}=1$) behaves at large $\eta$ as 
\begin{equation}
\label{h-fall-off-AdS3}
    h_{ij}(\eta) = e^{2 \eta} h_{ij}^{(0)} + \mathcal O(1) \ . 
\end{equation}

\subsection{Regularity of the supergravity solution} 
\label{sec:regularity-ald}

In this section, we discuss the regularity of the asymptotically linear dilaton ($\ALD_3$) solutions \eqref{eq:ald-sols} for different values of the parameters. We have restricted the AdS$_3$ solutions \eqref{eq:ads3-main-metric} to masses $8M \geq -1$, in order to avoid objects with negative tension. For the same reason, the asymptotically linear dilaton geometries \eqref{eq:main-metric} are only well-defined for $8M \geq -1$. There is another condition on solutions with negative mass: in order for the square-root entering the dilaton solution \eqref{eq:main-dilaton} to be positive, we require the positive deformation parameter $R_\phi$ to obey
\begin{equation}
    R_\phi^2 \geq - 8 M \ap n_5 \ .  
\end{equation}
Let us now analyze the properties of the solutions \eqref{eq:ald-sols} for different values of $M$. 

\paragraph{$\boldsymbol{8M=-1}$: global $\boldsymbol{\ALD}_{\boldsymbol 3}$ and thermal $\boldsymbol{\ALD}_{\boldsymbol 3}$.} Similar to the $\AdS_3$ case analyzed in the previous section, the solution is regular everywhere and defined for any $r \geq 0$. We will again be interested in the geometry with a compactified Euclidean time. The metric interpolates from thermal $\AdS_3$ in the interior to $\ALD_3$  asymptotics. No condition is imposed on the Euclidean time periodicity.

\paragraph{$\boldsymbol{-1 < 8M < 0}$: $\boldsymbol{\ALD}_{\boldsymbol 3}$ conical defect geometries.} When $8M = - \frac{1}{n^2}$ with $n \in \mathbb Z$, the compact coordinate $\phi$ acquires again a deficit angle $\phi \sim \phi + \frac{2 \pi}{n}$ and we are again interested in compact Euclidean time. No condition is imposed on the periodicity and the geometry is defined for any $r > 0$.

\paragraph{$\boldsymbol{M=0}$: $\boldsymbol{\ALD}_{\boldsymbol 3}$ cusp geometry.} As in the AdS$_3$ cusp case reviewed above, there is a naked singularity at $r=0$. One is again free to compactify the Euclidean time coordinate. This case is special because the square-root structure  in the dilaton profile \eqref{eq:main-dilaton} is not visible, unlike cases with a non-zero $M$ which provides a new mass scale.

\paragraph{$\boldsymbol{M > 0}$: $\boldsymbol{\ALD}_{\boldsymbol 3}$ black hole geometry.} For $M>0$, the geometry \eqref{eq:main-metric} shares various features with the BTZ black hole. There is a horizon located at
\begin{equation} 
\label{eq:rp-ald}
     r = r_+ = \sqrt{8 M \ap n_5} \ .
\end{equation}
In order for the geometry to be regular and to cap off smoothly at $r=r_+$, the Euclidean time coordinate $t$ needs to be compact with a periodicity related to the mass:\footnote{As expected, eq.~\eqref{deformed-BTZ-beta} reduces to \eqref{BTZ-beta} in the decoupling limit $R_\phi \to \infty$. }
\begin{equation}
    \beta_{{\rm BH}} = 2 \pi \sqrt{\frac{\ap n_5}{8 M} \biggl(1 + \frac{8 M \ap n_5}{R_\phi^2} \biggr)} \ .  
    \label{deformed-BTZ-beta}
\end{equation}

\section{Holographic renormalization and the variational principle}
\label{sec:logic}

Unless suitable boundary terms are included, the AdS$_3$ on-shell action \eqref{AdS3-bulk-action} diverges. 
The underlying reason is that the manifold $\mathcal{M}_3=\text{AdS}_3$ is non-compact and the integral over the radial coordinate $\eta$ diverges. The same issue occurs when considering ALD geometries discussed in Section~\ref{sec:ald} \cite{Marolf:2007ys, Mann:2009id}. 

In light of AdS/CFT, there is a natural way to extract the finite value of the gravity on-shell action following the usual renormalization procedure in quantum field theory. This procedure is known as `holographic renormalization'; for a review, see~\cite{Skenderis:2002wp}. The UV divergences in the dual theory, thought of as living at the boundary, are associated with IR divergences in the bulk coming from the large $\eta$ integration region. The introduction of a UV cutoff in the dual quantum field theory corresponds, via holography, to the introduction of an infrared radial cutoff at $\eta=\eta_c$. One then considers all possible boundary `counter-terms' consistent with the symmetries of the problem with the hope of rendering the divergent bulk action finite in the large $\eta_c$ limit. 

While this approach was initially well motivated from the usual renormalization procedure in quantum field theory, the choice of boundary terms has a direct motivation within semi-classical gravity. That motivation is partially provided by sharpening the variational principle. On spacetimes with the desired asymptotic behavior, the possible boundary terms are constrained by requiring that the variation of the action vanish identically when the equations of motion are imposed~\cite{Regge:1974zd, Banados:1992wn, Papadimitriou:2005ii, Mann:2005yr, Mann:2009id}. 

We will follow this approach. In the examples studied here, it turns out that the on-shell action is finite as a consequence of this requirement that the action be stationary. We will see concrete examples of this mechanism in Section~\ref{sec:on-shell}. 
Note that this approach still leaves some freedom to add boundary terms that are finite. 
In Section~\ref{sec:on-shell}, we will fix this arbitrariness by a physical ansatz and by requiring that the on-shell action for the dilaton geometries \eqref{eq:ald-sols} smoothly reduces to the known AdS$_3$ result in the limit $R_\phi \to \infty$.  

Let us describe how the action principle can be used to identify the boundary terms necessary to regularize the bulk action for a spacetime $\mathcal M_d$. Assume the metric can be written in normal coordinates \eqref{eq:ald-normal} and that $\mathcal M_d$ is either a compact manifold with boundary at $\eta = \eta_c$, or that it admits an asymptotic foliation into sub-manifolds $\partial \mathcal M_d^{\eta_c}$, identified by different values of $\eta = \eta_c$.

Consider a bulk action, $S_{\text{bulk}}$, capturing the dynamics of a set of fields $\{ \mathcal{F}(\eta,\mathbf{x}) \}$ on the $d$-dimensional spacetime $\mathcal{M}_d$,
\begin{equation}
    S_{\text{bulk}} = \int_{\mathcal{M}_d} \text d\eta \, \text d^{d-1}\mathbf{x}  \, \mathcal L_\text{bulk}(\{\mathcal{F}(\eta,\mathbf{x})\}) \ ,
    \label{bulk-action-abstract}
\end{equation}
where $\mathcal L_\text{bulk}$ is the Lagrangian density and $\eta$ is a distinguished normal coordinate; see eq.~\eqref{eq:ald-normal}. For the example of AdS$_3$, which we are going to discuss in detail below, we have $d=3$, $\mathbf{x} = (t, \phi)$, the collection of fields is given by $\{ \mathcal{F}(\eta,\mathbf{x}) \} = \{ \hat g_{\mu\nu}(\eta, t , \phi)\}$. The action $S_{\text{bulk}}$ is given by eq.~\eqref{AdS3-bulk-action}. For the asymptotically linear dilaton solutions~\eqref{eq:ald-sols}, again $d=3$, $\mathbf{x} = (t, \phi)$, but the collection of fields includes the dilaton, $\{ \mathcal{F}(\eta,\mathbf{x}) \} = \{ \hat g_{\mu\nu}, \Phi \}$ and the bulk action is eq.~\eqref{S3-final-Einstein}. 

\subsection{Compact manifold with boundary}
\label{sec:fintite-volume-with-boundary}

Let us first consider the case of a compact manifold $\mathcal M_d$ with boundary $\partial \mathcal M_d$ at $\eta = \eta_c$. Here $\eta_c$ is a fixed value depending on the parametrization of the manifold. The boundary terms one may consider take the form 
\begin{equation}
\label{S-bound}
    S_{\text{bound}} = \int_{\partial \mathcal{M}_d} \text d^{d-1} \mathbf{x} \, \mathcal L_{\text{bound}} (\{ \mathcal{F}(\eta_c, \mathbf{x}) \}) \,.
\end{equation}
Our goal is to identify the necessary boundary terms $\mathcal L_{\text{bound}}$ needed for the variational principle with action
\begin{equation}
\label{Stot-def}
    S_{\text{tot}} = S_{\text{bulk}} + S_{\text{bound}}
\end{equation}
to be well-defined. 

Let us impose Dirichlet boundary conditions on the fields $\mathcal{F}(\eta,\textbf{x})$ at $\eta=\eta_c$, 
\begin{equation} \label{eq:dirichlet-compact}
    \delta \mathcal{F}(\eta_c,\mathbf{x}) = 0 \ , 
\end{equation}
and consider the variation of the bulk action \eqref{bulk-action-abstract},\footnote{We are assuming that $S_{\text{bulk}}$ contains at most two derivatives.} 
\begin{equation}
\label{bulk-variation-abstract}
    \delta S_{\text{bulk}} = \sum_{\mathcal{F}} \int_{\mathcal{M}_d} \text d \eta \, \text d^{d-1} \mathbf{x} \, E_{\mathcal{F}} \, \delta \mathcal{F}(\eta,\mathbf{x}) + \sum_{\mathcal{F}}\int_{\partial \mathcal{M}_d} \text d^{d-1} \mathbf{x} \, B_{\mathcal{F}} \,  \partial_\eta \delta\mathcal{F}(\eta,\mathbf{x})\Big|_{\eta=\eta_c} \ ,
\end{equation}
where no boundary term of the form $\delta \mathcal F(\eta_c, \mathbf x)$ appears because of the boundary condition \eqref{eq:dirichlet-compact}. One would expect that the variational principle, 
\begin{equation}
\label{variational-principal-bk}
    \delta S_{\text{bulk}} = 0\,,
\end{equation}
implies the equations of motion, 
\begin{equation} 
\label{eq:compact-eom}
    E_{\mathcal{F}} = 0 \ , 
\end{equation}
for the fields $\mathcal{F}(\eta,\mathbf{x})$. However, if
\begin{equation}
     B_{\mathcal{F}} \neq 0 \ , 
\end{equation}
the second term in \eqref{bulk-variation-abstract} is non-zero and eq.~\eqref{variational-principal-bk} does not imply the equations of motion~\eqref{eq:compact-eom}. For instance, when the theory under consideration is pure gravity on a compact manifold with boundary, the bulk action \eqref{bulk-action-abstract} is the Einstein-Hilbert action and the additional boundary term \eqref{S-bound} amounts to the Gibbons-Hawking-York term \cite{York:1972sj, Gibbons:1976ue}
\begin{equation} \label{eq:gh-term-ads3}
    S_{\text{GHY}} = -\frac{1}{\kappa_d^2} \int_{\partial \mathcal{M}_d} \text d^{d-1} \mathbf{x} \, \sqrt{h} \, \text{Tr} K \ ,
\end{equation}
expressed in terms of the induced metric $h$ and the extrinsic curvature $K$. For a discussion about deriving the GHY term from string theory, see~\cite{Ahmadain:2024uyo, Ahmadain:2024uom}. The variational principle, 
\begin{equation}
    \delta S_{\text{tot}} = \delta S_{\text{bulk}} + \delta S_{\text{GHY}} = 0 \,,
\end{equation}
then correctly implies the Einstein equations of motion with our chosen Dirichlet boundary conditions. 

\subsection{Manifold foliated by spatial slices}

Let us alternately assume there exists an asymptotic foliation of $\mathcal{M}_d$ into sub-manifolds $\partial \mathcal{M}_d^{\eta_c}$, identified by constant values of $\eta= \eta_c$, and let $\bf{x}$ denote coordinates on $\partial \mathcal{M}_d^{\eta_c}$. Since we are analyzing Euclidean backgrounds, this guarantees that normal coordinates of the form \eqref{eq:ald-normal} can be chosen. 
The bulk action integral \eqref{bulk-action-abstract} can then be infra-red regularized 
\begin{equation} \label{eq:reg-s-bk}
    S_{\text{bulk}}(\eta_c) = \int^{\eta_c}_0 \text d\eta \int_{\partial \mathcal M_d^\eta} \text d^{d-1}\mathbf{x}  \, \mathcal L_\text{bulk}(\{\mathcal{F}(\eta,\mathbf{x})\}) \ .  
\end{equation}
As mentioned earlier, the action is not generally stationary on backgrounds with the desired asymptotic behavior and divergences arise from the limit $\eta_c\to \infty$. One hopes to be able to cure both issues by adding (finitely many) boundary terms. For fixed $\eta_c$, these terms take the form 
\begin{equation}
\label{S-bound-foliation}
    S_{\text{bound}}(\eta_c) = \int_{\partial \mathcal{M}_d^{\eta_c}} \text d^{d-1}\mathbf{x} \, \mathcal L_{\text{bound}} (\{ \mathcal{F}(\eta_c, \mathbf{x}) \}) \,,
\end{equation}
with the total action given by the sum of bulk and boundary contributions, 
\begin{equation} \label{eq:tot-action-non-compact}
    S_{\text{tot}}(\eta_c) = S_{\text{bulk}}(\eta_c) + S_{\text{bound}}(\eta_c) \,. 
\end{equation}
We then define the on-shell action as: 
\begin{equation}
\label{S-on-shell}
    S_{\text{on-shell}} = \lim_{\eta_c \to \infty} S_{\text{tot}}(\eta_c) \ .  
\end{equation}
For the cases of interest to us, we will see below that this procedure guarantees that the action \eqref{S-on-shell} is finite and agrees with expectations from thermodynamics.  

\paragraph{Asymptotic fall-off and boundary conditions.}  For the various fields $\mathcal F(\eta, \mathbf x)$ entering the bulk action, we assume the asymptotic fall-off at large $\eta$ 
\begin{equation} 
\label{eq:f-bdy-conditions}
    \mathcal{F}(\eta,\mathbf{x}) = f_{0}(\eta) \mathcal{F}^{(0)}(\mathbf{x}) + f_{1}(\eta) \mathcal{F}^{(1)}(\mathbf{x}) + f_{2}(\eta) \mathcal{F}^{(2)}(\mathbf{x}) + \ldots \ ,
\end{equation}
where $f_{0}(\eta)$ specifies the leading behavior, while $f_j(\eta)$ with $j \geq 1$ label subleading terms. 
We will illustrate this specifically for asymptotically AdS$_3$ and \ALD$_3$ geometries in the subsequent sections. For pure gravity in the AdS$_3$ case, the transverse components of the metric $h_{ij}$ behave like
\begin{equation} \label{eq:bdy-cdts-ads3}
    f_{0}(\eta) = e^{2\eta} \ , \quad f_1(\eta)=1 \ , \quad f_{j\geq 2}(\eta)\sim \mathcal{O}(e^{-\eta}) \ , \quad \mathcal{F}^{(0)}(\mathbf{x}) = h_{ij}^{(0)} \ , 
\end{equation}
where $h_{ij}^{(0)}$ does not depend on $\mathbf x$. Equivalently 
\begin{equation}
    h_{ij} = e^{2 \eta}h_{ij}^{(0)} + h_{ij}^{(1)} + \ldots\,.
\end{equation}
Dirichlet boundary condition are now implemented by requiring that $\mathcal{F}^{(0)}(\mathbf{x})$ is kept fixed while the variation $\delta \mathcal{F}^{(n)}(\mathbf{x})$ for $n>0$ need not vanish,\footnote{In principle, one could consistently impose that a few $\mathcal{F}^{(n)}(\mathbf{x})$ are also fixed on a particular set of solutions, but one should make sure that this is consistent with the variation of the action. For example, in the case of asymptotically AdS$_3$, one can set the coefficient of $e^{\eta}$ to zero, as it is already done in \eqref{eq:bdy-cdts-ads3}. \label{footnote:some-zero-bdy}}
\begin{equation} 
\label{eq:non-compact-bdy-conditions}
    \delta \mathcal{F}^{(0)}(\mathbf{x})=0 \ , \qquad \delta \mathcal F^{(n)}(\mathbf{x}) \neq 0 \ \ \text{ for } \ \ n>0 \ .
\end{equation}

\paragraph{Boundary terms.} The next step is to identify boundary terms so that the action principle is well-defined. After introducing a cutoff at $\eta=\eta_c$, the variation of the regularized bulk action~\eqref{eq:reg-s-bk} generically takes the form
\begin{multline} 
\label{eq:variation-bdy-terms}
    \delta S_{\text{bulk}}(\eta_c) =  \sum_{\mathcal{F}}  \int^{\eta_c}_0 \text d\eta \int_{\partial \mathcal{M}_d^{\eta_c}}  \, \text d^{d-1} \mathbf{x} \, E_{\mathcal{F}} \, \delta \mathcal{F}(\eta,\mathbf{x}) + \sum_{\mathcal{F}}\int_{\partial \mathcal{M}_d^{\eta_c}} \text d^{d-1} \mathbf{x} \, A_{\mathcal{F}} \, \delta \mathcal{F}(\eta,\mathbf{x})\Big|_{\eta=\eta_c} \\
    + \sum_{\mathcal{F}}\int_{\partial \mathcal{M}_d^{\eta_c}} \text d^{d-1} \mathbf{x} \, B_{\mathcal{F}} \,  \partial_\eta \delta\mathcal{F}(\eta,\mathbf{x})\Big|_{\eta=\eta_c}\ ,
\end{multline}
where $A_{\mathcal{F}}$ and $B_{\mathcal{F}}$ are functionals of the fields $\{ \mathcal F(\eta, \mathbf x)\}$. Notice that compared to the case of a compact manifold with boundary, discussed in Section~\ref{sec:fintite-volume-with-boundary}, eq.~\eqref{eq:variation-bdy-terms} contains extra terms proportional to $ A_{\mathcal F} \, \delta \mathcal F(\eta, \mathbf x)$ when compared with eq.~\eqref{bulk-variation-abstract}. This is a consequence of the boundary conditions \eqref{eq:non-compact-bdy-conditions}, which generically do not imply $\delta \mathcal{F}(\eta,\mathbf{x})\Big|_{\eta=\eta_c} = 0$. We will see explicit examples of these extra terms for both asymptotically AdS$_3$ and \ALD$_3$ geometries. In both cases, these terms do not vanish in the large $\eta_c$ limit. 

As a consequence, the variational principle $\delta S_{\text{bulk}}(\eta_c)=0$ again does not imply the equations of motion 
\begin{equation} \label{eq:non-compact-eom}
    E_{\mathcal{F}} = 0 \ ,
\end{equation}
and additional boundary terms should be added to the action. We will see in concrete examples below that the resolution is the one we discussed around eq.~\eqref{S-bound-foliation}. Namely, adding suitable boundary terms to the bulk action so that the action principle is restored for the total action \eqref{eq:tot-action-non-compact} and the on-shell action \eqref{S-on-shell} is finite. 

\subsection[Asymptotically AdS$_3$ geometries]{Asymptotically AdS$_{\boldsymbol 3}$ geometries} \label{sec:bdy-terms-ads3-identified}

As a warm up for the ALD cases, let us make the formalism concrete in the well-studied case of pure gravity in the asymptotically AdS$_3$ geometries~\eqref{eq:ads3-main-metric}.\footnote{Every asymptotically AdS$_3$ geometry is also locally AdS$_3$ and hence of the form \eqref{eq:ads3-main-metric}.} From eq.~\eqref{h-fall-off-AdS3} we define asymptotically AdS$_3$ backgrounds by the Fefferman-Graham asymptotic form,
\begin{equation}
    h_{ij}(\eta) = e^{2\eta} \, h^{(0)}_{ij}(\mathbf x) + h^{(1)}_{ij}(\mathbf x) + \mathcal{O}(e^{-\eta}) \ ,
    \label{AdS3-eta-falloff}
\end{equation}
where $h^{(0)}_{ij}$ is kept fixed, i.e.
\begin{equation}
    \delta h_{ij}(\eta) = \delta h^{(1)}_{ij} + \mathcal{O}(e^{-\eta}) \ .
\end{equation}
Notice that no term proportional to $e^{\eta}$ enters \eqref{AdS3-eta-falloff}. 

Consider the variation of the action \eqref{AdS3-bulk-action}. After introducing a cutoff at $\eta = \eta_c$, the bulk Einstein-Hilbert action on $\mathcal{M}_3^{\eta_c}$ can be written as 
\cite{Kraus:2006wn}
\begin{multline} 
\label{eq:EH-ads3}
	S_{\text{bulk}}(\eta_c) = - \frac{1}{16\pi G_3} \int_{\mathcal{M}_3^{\eta_c}} \text d \eta \, \text d^2 \mathbf{x} \, \sqrt{h} \big[ R^{(2)} + (\text{Tr} K)^2 - \text{Tr} K^2 \big] \\
    + \frac{1}{8\pi G_3} \int_{\partial \mathcal{M}_3^{\eta_c}} \text d^2 \mathbf{x} \, \sqrt{h} \, \text{Tr} K \ ,
\end{multline}
where $R^{(2)}$ is the two-dimensional Ricci scalar defined in terms of the boundary metric $h_{ij}$, $h$ is the determinant of $h_{ij}$, the extrinsic curvature $K_{ij}$ on $\partial \mathcal{M}_3^{\eta_c}$ reads, 
\begin{equation} \label{eq:extrinsic-curvature-no-compact}
	K_{ij} = \tfrac{1}{2} \partial_{\eta} h_{ij} \Big|_{\eta=\eta_c}  \ , 
\end{equation}
and
\begin{equation} \label{eq:def-compact-k}
	\text{Tr} K = h^{ij} K_{ij} \ , \quad \text{Tr} K^2 = K_{ij} K^{ij} \ . 
\end{equation} 
This rewriting of the Einstein-Hilbert action is special to three dimensions. 
One can see that the variation of the bulk action \eqref{eq:EH-ads3} contains boundary terms of the form \eqref{eq:variation-bdy-terms}. Unlike the case of a compact manifold with boundary, discussed in Section~\ref{sec:fintite-volume-with-boundary}, adding the Gibbons-Hawking-York term 
\begin{equation} 
\label{eq:gh-term-ads3-non-compact}
    S_{\text{GHY}}(\eta_c) = - \frac{1}{8\pi G_3} \int_{\partial \mathcal{M}_3^{\eta_c}} \text d^2 \mathbf{x} \, \sqrt{h} \, \text{Tr} K\,, 
\end{equation}
is not sufficient. The second term in the first line of \eqref{eq:variation-bdy-terms} is non-zero. In fact, varying $S_{\text{bulk}}(\eta_c)+S_{\text{GHY}}(\eta_c)$ and imposing the fall-off conditions \eqref{AdS3-eta-falloff} gives
\begin{multline}
\label{delta-Sbulk+SGH}
	\delta S_{\text{bulk}}(\eta_c)+ \delta S_{\text{GHY}}(\eta_c) =  \int^{\eta_c}_0 \text d \eta \int_{\partial \mathcal{M}_3^{\eta_c}} \text d^2 \mathbf{x} \, E_g^{\mu \nu} \, \delta g_{\mu \nu}(\eta,\mathbf x) \\
    - \frac{1}{16\pi G_3} \int_{\partial \mathcal{M}_3^{\eta_c}} \text d^2 \mathbf{x} \, \delta h^{(2)}_{ij} h^{(0)ij} \sqrt{h^{(0)}} + \mathcal{O}(e^{-\eta_c}) \ . 
\end{multline}
This shows that with the fall-off conditions \eqref{AdS3-eta-falloff}, the variational principle does not imply the equations of motion
\begin{equation}
    (E_{g})_{\mu\nu} = \hat{R}_{\mu\nu} - \frac{1}{2} \hat{R} \, \hat g_{\mu\nu} - \hat g_{\mu\nu} = 0 \ .  
    \label{AdS3-EoM}
\end{equation}
As discussed above, the resolution is to consider additional boundary terms. In particular, eq.~\eqref{delta-Sbulk+SGH} suggests defining
\begin{equation} 
\label{eq:ads3-ct}
	S_{\text{bound}}(\eta_c)= S_{\text{GHY}}(\eta_c) + \frac{1}{8\pi G_3} \int_{\partial \mathcal{M}_3^{\eta_c}} \text d^2 \mathbf{x} \, \sqrt{h} \ .
\end{equation}
One can see that the variation of the total action
\begin{equation}
    S_{\text{tot}}(\eta_c) = S_{\text{bulk}}(\eta_c) + S_{\text{bound}}(\eta_c) \ , 
\end{equation}
with the boundary action defined as in eq.~\eqref{eq:ads3-ct}, reads 
\begin{equation}
    \delta S_{\text{tot}}(\eta_c) = \int^{\eta_c}_0 \text d \eta \int_{\partial \mathcal{M}_3^{\eta_c}} \text d^2 \mathbf{x} \, E_g^{\mu \nu} \, \delta \hat g_{\mu \nu}(\eta,\mathbf x)
\end{equation}
and hence correctly implies the equations of motion~\eqref{AdS3-EoM}. We emphasize that the boundary term entering eq.~\eqref{eq:ads3-ct} is fixed for \textit{any} asymptotically AdS$_3$ solution. The counter-term \eqref{eq:ads3-ct} that we have obtained is the standard counter-term used in the calculation of the on-shell action for asymptotically AdS$_3$ geometries, see e.g.\ \cite{Carlip:1994gc,Dei:2024sct}.

\subsection{Asymptotically linear dilaton geometries} 
\label{sec:ald-on-shell-comptutation}

The goal of this section is to identify the necessary boundary terms for the dimensionally reduced action \eqref{S3-final-Einstein} and the associated geometries \eqref{eq:ald-sols}. 

\paragraph{Asymptotic fall-off and boundary conditions.} We now need to impose  boundary conditions for the metric and dilaton. It is convenient to again use the normal-coordinate system \eqref{eq:ald-normal}. Eqs.~\eqref{Phi-falloff} and \eqref{h-falloff} suggest defining \ALD$_3$ geometries by the fall-off conditions
\begin{subequations} 
\label{eq:ald-bdy-conditions}
\begin{equation} 
\label{eq:ald-metric-set-f}
	h_{ij} = \eta^2 h^{(0)}_{ij} + \eta \, h^{(1)}_{ij} + \mathcal{O}(1) \ ,
\end{equation}
\begin{equation} 
\label{eq:phi-set-f}
	\Phi = -\frac{1}{2} \log{\eta} + \Phi^{(0)} + \eta^{-1} \Phi^{(1)} + \mathcal{O}(\eta^{-2}) \ ,
\end{equation}  
\end{subequations}
where $\Phi^{(0)}$ is constant and in our case reads:  
\begin{equation}
    \Phi^{(0)} = \frac{1}{2}\log\Bigl(\frac{\sqrt{\ap n_5}}{2}\Bigr) \ . 
\end{equation}
Eqs.~\eqref{eq:ald-bdy-conditions} agree with the standard examples in the literature, see e.g.~\cite{Mann:2009id,Chang:2023kkq}. For these asymptotically linear dilaton geometries,
boundary conditions are imposed by requiring\footnote{Notice that fixing $\delta h_{ij}^{(0)} = 0$ does not necessarily imply $\delta h_{ij}^{(1)} = 0$, since $h^{(1)}_{ij}$ is not uniquely fixed in terms of $h^{(0)}_{ij}$.}
\begin{align}
    \delta h_{ij} &= \eta \, \delta h^{(1)}_{ij} + \mathcal{O}(\eta^{-2}) \ , \\
    \delta \Phi &=  \eta^{-1} \delta \Phi^{(1)}  + \mathcal{O}(\eta^{-2}) \ . 
\end{align}
To identify the needed boundary terms, we begin by considering the variation of the bulk action \eqref{S3-final-Einstein} together with the Gibbons-Hawking-York term~\eqref{eq:gh-term-ads3-non-compact}, 
\begin{multline} 
\label{eq:bdy-i0}
	\delta S_{\text{bulk}}(\eta_c) + \delta S_{\text{GHY}}(\eta_c) = \int^{\eta_c}_0 \text d \eta \int_{\partial \mathcal{M}_3^{\eta_c}} \text d^2 \mathbf{x} \, \Bigl(  E_g^{\mu \nu} \, \delta \hat g_{\mu \nu}(\eta,\mathbf x) + E_\Phi \, \delta \Phi(\eta)  \Bigr) \\
    - \frac{1}{2\kappa_3^2} \int_{\partial \mathcal{M}_3^{\eta_c}} \text d^2 \mathbf{x} \, \sqrt{h} \, \delta h_{ij} (h^{ij} \text{Tr} K - K^{ij}) + \frac{4}{\kappa_3^2} \int_{\partial \mathcal{M}_3^{\eta_c}} \text d^2 \mathbf{x} \, \sqrt{h} \, \partial_{\eta} \Phi \, \delta \Phi \ ,
\end{multline}
where we again use normal coordinates \eqref{eq:ald-normal}; see eqs.~\eqref{eq:extrinsic-curvature-no-compact} and \eqref{eq:def-compact-k}. Making use of \eqref{eq:ald-bdy-conditions}, we see
\begin{multline} 
\label{eq:unwanted-var}
	\delta S_{\text{bulk}}(\eta_c) + \delta S_{\text{GHY}}(\eta_c) = \int^{\eta_c}_0 \text d \eta \int_{\partial \mathcal{M}_3^{\eta_c}} \text d^2 \mathbf{x} \, \Bigl(  E_g^{\mu \nu} \, \delta \hat g_{\mu \nu}(\eta,\mathbf x) + E_\Phi \, \delta \Phi(\eta)  \Bigr)\\
    -\frac{1}{2\kappa_3^2} \int_{\partial \mathcal{M}_3^{\eta_c}} \text d^2 \mathbf{x} \, \delta h^{(1)}_{ij} h^{(0)ij} \sqrt{h^{(0)}} - \frac{2}{\kappa_3^2} \int_{\partial \mathcal{M}_3^{\eta_c}} \text d^2 \mathbf{x} \, \sqrt{h^{(0)}} \delta \Phi^{(1)} + \mathcal{O}(\eta_c^{-1}) \ .
\end{multline}
This shows that additional boundary terms are again necessary. 

Let us formulate an ansatz for the possible boundary terms.
Since we are considering a bulk action that includes at most $2$ derivatives, the variation in eq.~\eqref{eq:bdy-i0} contains terms with at most one derivative. We therefore restrict to boundary terms with at most $2$ derivatives. At this point we need to impose a physical assumption. The Gibbons-Hawking-York term is constructed from the boundary metric and the normal derivative $n^\mu \partial_\mu = \partial_\eta$, which is why the extrinsic curvature can appear and not just diffeomorphism invariant couplings intrinsic to the boundary. This is needed to cancel the bulk variation, which for pure gravity contains metric times normal derivative of the variation of the metric, integrated over the boundary coordinates; see \C{eq:EH-ads3}. On the other hand, the bulk variation of a scalar field like the dilaton does not produce such a term and one can consistently impose Dirichlet boundary conditions. So, a priori, there is no need to include boundary couplings beyond $S_{\text{GHY}}$ which involve normal derivatives. We will take this as part of our definition of Dirichlet boundary conditions. This same view point was taken in the earlier work on boundary terms for ALD spacetimes~\cite{Marolf:2007ys, Mann:2009id}.\footnote{We would like to thank Bob McNees for very helpful correspondence on this point.} 

With the physical motivation stated above, we restrict any additional boundary terms beyond the GHY term to invariants under two-dimensional diffeomorphisms acting on $\partial \mathcal M_3^{\eta_c}$. We therefore consider the ansatz
\begin{equation}
\label{ansatz-Sbound}
    S_{\text{bound}}(\eta_c) = S_{\text{GHY}}(\eta_c) + \int_{\partial \mathcal{M}_3^{\eta_c}} \text d^2 \mathbf{x} \, \sqrt{h} \, P(e^{2\Phi}) + \int_{\partial \mathcal{M}_3^{\eta_c}} \text d^2 \mathbf{x} \, \sqrt{h} \, R^{(2)} Q(e^{2\Phi}) \ .
\end{equation}
Let us explain the various ingredients entering \eqref{ansatz-Sbound}. The only diffeomorphism invariants constructed from the metric $h_{ij}$ are $\sqrt{h}$ and $\sqrt{h} \, R^{(2)}$. The latter coupling is topological unless there is a non-vanishing coupling to the scalar field. The dilaton $\Phi \sim \log \eta$ for large $\eta$ but there is no bulk variation that can generate compensating log dependence; hence we choose to restrict the functions $P$ and $Q$ to analytic functions of $e^{2\Phi}$. The variation of the $\sqrt{h} \, R^{(2)}$ in \eqref{ansatz-Sbound} includes the Einstein tensor, which vanishes identically in two dimensions, as well as a total derivative term. Integrating by parts then gives terms proportional to boundary derivatives of $Q(e^{2\Phi})$. However, there are no corresponding terms from the bulk variation \eqref{eq:unwanted-var} that also involve derivatives in the boundary directions.  We can therefore discard the second term of \eqref{ansatz-Sbound} and restrict our ansatz to the form: 
\begin{equation} 
\label{eq:ald-counter-terms}
    S_{\text{bound}}(\eta_c) = S_{\text{GHY}}(\eta_c) + \int_{\partial \mathcal{M}_3^{\eta_c}} \text d^2 \mathbf{x} \, \sqrt{h} \, P(e^{2\Phi}) \ .
\end{equation}
The corresponding variation reads 
\begin{equation}
    \delta S_{\text{bound}}(\eta_c) - \delta S_{\text{GHY}}(\eta_c) = \int_{\partial \mathcal{M}_3^{\eta_c}} \text d^2 \mathbf{x} \, \sqrt{h} \left[\frac{1}{2} h^{ij}\delta h_{ij} P(e^{2\Phi}) + 2 P^{\prime}(e^{2\Phi}) e^{2\Phi} \delta\Phi \right] \ .
\end{equation}
Using the large $\eta_c$ expansion in eq.~\eqref{eq:ald-bdy-conditions} gives the leading behavior,
\begin{multline} 
\label{eq:f-variation-generic}
    \delta S_{\text{bound}}(\eta_c) - \delta S_{\text{GHY}}(\eta_c) = \int_{\partial \mathcal{M}_3^{\eta_c}} \text d^2 \mathbf{x} \, \sqrt{h^{(0)}} \Big[\frac{\eta_c}{2} \,  h^{(0)ij} \, \delta h^{(1)}_{ij} P\left(\frac{\sqrt{\ap n_5}}{2 \, \eta_c}\right) \\
    + 2 P^{\prime}\left(\frac{\sqrt{\ap n_5}}{2 \eta_c}\right) \frac{\sqrt{\ap n_5}}{2} \, \delta \Phi^{(1)} \Big] + \ldots \ .
\end{multline}
In order to understand the possible boundary terms, let us expand
\begin{equation}
    P(x) = \frac{2}{\kappa_3^2\sqrt{\ap n_5}} \sum_{n\geq 0} P_n x^n \ ,
\end{equation}
near $x=0$. Comparing eq.~\eqref{eq:f-variation-generic} with \eqref{eq:unwanted-var}, we see that $P_0=0$. Moreover, the terms $P_n \, x^n$ with $n\geq 2$ do not contribute to $\delta S_{\text{bound}}(\eta_c)$ in the large $\eta_c$ limit. However, the term $P_2 \, x^2$ contributes a finite piece to the on-shell action. This is because at large $\eta_c$, the leading term in the on-shell action is
\begin{equation}
    S_{\text{bound}}(\eta_c) = S_{\text{GHY}}(\eta_c) + \int_{\partial \mathcal{M}_3^{\eta_c}} \text d^2 \mathbf{x} \, \sqrt{h^{(0)}} \, \eta_c^2 \, P\left(\frac{\sqrt{\ap n_5}}{2\eta_c}\right) + \ldots \ .
\end{equation}
Therefore, we consider
\begin{equation}
    P(x) = \frac{2}{\kappa_3^2\sqrt{\ap n_5}} \left( P_1 x + P_2 x^2 \right) \ ,
\end{equation}
while neglecting terms of the form $P_n x^n$ with $n\geq 3$. In summary, we consider the boundary terms: 
\begin{equation} 
\label{eq:ald-bdy-terms-final}
    S_{\text{bound}}(\eta_c) = S_{\text{GHY}}(\eta_c)+ \frac{2 P_1}{\kappa_3^2\sqrt{\ap n_5}} \int_{\partial \mathcal{M}_3^{\eta_c}} \text d^2 \mathbf{x} \, \sqrt{h} \, e^{2\Phi} + \frac{2 P_2}{\kappa_3^2\sqrt{\ap n_5}} \int_{\partial \mathcal{M}_3^{\eta_c}} \text d^2 \mathbf{x} \, \sqrt{h} \, e^{4\Phi} \ .
\end{equation}
Using the variation in \eqref{eq:f-variation-generic} and imposing stationarity, 
\begin{equation}
    \delta S_{\text{tot}} = \delta S_{\text{bulk}} + \delta S_{\text{bound}} = 0 \ , 
\end{equation}
at large $\eta_c$ gives
\begin{equation}
    P_1 = 1 \ ,
\end{equation}
while $P_2$ remains unfixed. This is expected since the term proportional to $P_2$ in \eqref{eq:ald-bdy-terms-final} does not contribute to the variation of the action in the large $\eta_c$ limit. 
To fix the one-parameter freedom in the action specified by $P_2$, we will require that the on-shell action be smooth as $R_\phi \to \infty$. We will discuss this matching condition in the next section.

\section{On-shell action} 
\label{sec:on-shell}

In this section, we come to the core of the paper and compute the on-shell action for the ALD$_3$ backgrounds discussed in Section~\ref{sec:ald}. Before delving into the details of the ALD$_3$ calculation, let us briefly recap how the on-shell action is computed for asymptotically AdS$_3$ geometries.

\subsection[Asymptotically AdS$_{3}$ geometries]{Asymptotically AdS$_{\boldsymbol 3}$ geometries}
\label{sec:ads3-on-shell-calc}

We are now equipped with all the necessary ingredients to compute the on-shell action \eqref{S-on-shell}. The bulk contribution is given in eq.~\eqref{AdS3-bulk-action} and we identified the necessary boundary terms in eq.~\eqref{eq:ads3-ct}. We will compute the on-shell action for each of the locally AdS$_3$ geometries listed in Section~\ref{sec:AAdS}. At various stages in the computations below, we made use of the \texttt{Mathematica} differential geometry package by Matthew Headrick~\cite{headrick}.

\paragraph{$\boldsymbol{8M=-1}$: Thermal AdS$_{\boldsymbol 3}$.}
For thermal AdS$_3$, we obtain 
\begin{equation}
\label{S-on-shell-Th-AdS3}
    S_{\text{on-shell}} = -\frac{\bth}{8 G_3} \ ,
\end{equation}
in accord with the known result in the literature \cite{Banados:1992wn, Hawking:1982dh}. Using the Brown-Henneaux formula  for the central charge of the boundary theory~\cite{Brown:1986nw},
\begin{equation} \label{eq:brown-c}
    c = \frac{3 R_{\text{AdS}}}{2 G_3}  \ ,
\end{equation}
the on-shell action \eqref{S-on-shell-Th-AdS3} can be rewritten as 
\begin{equation}
    S_{\text{on-shell}} = -  \frac{\bth \, c}{12  R_{\text{AdS}}} \ ,
\end{equation}
which correctly reproduces the ground state energy of the boundary CFT, 
\begin{equation}
E = \, \partial_{\bth} S_{\text{on-shell}} = -\frac{c}{12 R_{\text{AdS}} }\ ,
\end{equation} 
where the radius of the boundary cylinder is $R_{\text{AdS}}$. One can also express the on-shell action as 
\begin{equation}
    S_{\text{on-shell}} = -\frac{c \pi \tau_2}{6} \ ,
\end{equation}
where 
\begin{equation}
    \tau_2 = \frac{\bth}{2\pi R_{\text{AdS}}} \ ,
\end{equation}
is the imaginary part of the modular parameter of the boundary torus.

\paragraph{$\boldsymbol{-1 < 8M < 0}$: Conical defect geometries.} This is very similar to the previous case except that now the periodicity of the angle $\phi$ is
\begin{equation}
    \phi \sim \phi + \frac{2\pi}{n} \ , \quad n \in \mathbb{Z} \ .
\end{equation}
We find the on-shell action
\begin{equation}
\label{conical-defect-S-on-shell}
    S_{\text{on-shell}} = -\frac{\beta_{\text{cd}}}{8 G_3 |n|} \ .
\end{equation}
Similar to thermal AdS$_3$, the on-shell action \eqref{conical-defect-S-on-shell} can be expressed in terms of the imaginary part of the modular parameter
\begin{equation} \label{eq:mod-tau2}
    \tau_2 = \frac{|n| \beta_{\text{cd}}}{2\pi R_{\text{AdS}}} \ ,
\end{equation}
as 
\begin{equation}
    S_{\text{on-shell}} = -\frac{c \pi \tau_2}{6 n^2} \ ,
\end{equation}
which agrees with the result in the literature~\cite{David:1999zb}. 

\paragraph{$\boldsymbol{M > 0}$: BTZ black hole.} In this case, we have checked that the value of the on-shell action equals\footnote{This can also be obtained by replacing $\bth$ with $\bth = \frac{4 \pi^2 R_{\text{AdS}}^2}{\beta_{\text{BTZ}}}$ in eq.~\eqref{S-on-shell-Th-AdS3}.} 
\begin{equation}
\label{S-on-shell-BTZ}
    S_{\text{on-shell}} = - \frac{\pi^2 R_{\text{AdS}}^2}{2 \beta_{\text{BTZ}} \, G_3}\ ,
\end{equation}
where we have used eq.~\eqref{BTZ-beta}. This is in agreement with expectations from thermodynamics: we can reproduce the Bekenstein-Hawking entropy,  
\begin{equation} \label{eq:entropy-ads3}
    \mathcal S = \beta_{\text{BTZ}} E - \beta_{\text{BTZ}} F =  \beta_{\text{BTZ}} \frac{\partial S_{\text{on-shell}}}{\partial \beta_{\text{BTZ}}} -S_{\text{on-shell}} = \frac{2 \pi r_+}{4 G_3} \ ,
\end{equation}
where $2\pi r_+$ is the black hole horizon area and we related the energy $E$ and the free energy $F$ to the on-shell action via \cite{Hawking:1982dh}
\begin{equation}
    E = \frac{\partial S_{\text{on-shell}}}{\partial \beta_{\text{BTZ}}} \ , \qquad  S_{\text{on-shell}} = \beta_{\text{BTZ}} \, F \ .
\end{equation}

\paragraph{$\boldsymbol{M=0}$: cusp geometry.} Introducing a cut-off at $r=\varepsilon$, for small $\varepsilon > 0$ to screen the naked singularity at $r=0$, we find that the on-shell action vanishes in the limit $\varepsilon\to 0$:
\begin{equation} \label{eq:on-shell-cusp}
    S_{\text{on-shell}} = 0 \ .
\end{equation}
This regulating procedure at $r=\varepsilon$ was already discussed in \cite{Dei:2024sct} and here we reproduce the same result. 
Since the cusp geometry is dual to the Ramond ground state of the boundary CFT, this is the expected result.

\paragraph{Spacetime mass.} For any $M \geq - \frac{1}{8}$, the spacetime mass can be derived from the on-shell action using,
\begin{equation}
     E = \frac{\partial S_{\text{on-shell}}}{\partial \beta}= \frac{M}{G_3} \ ,
\end{equation}
where $\beta$ denotes the periodicity of Euclidean time for each geometry under consideration; for example, $\beta = \beta_{\text{th}}$ for thermal AdS$_3$, $\beta = \beta_{\text{BTZ}}$ for the BTZ black hole and so on. 

\subsection{Asymptotically linear dilaton geometries}

Finally we discuss the computation of the on-shell action for the ALD$_3$ geometries described in Section~\ref{sec:regularity-ald}. Consider the action defined in eqs.~\eqref{eq:tot-action-non-compact} and \eqref{S-on-shell} with $S_{\text{bulk}}$ given by eq.~\eqref{S3-final-Einstein} and $S_{\text{bound}}$ by eq.~\eqref{eq:ald-bdy-terms-final} with $P_1=1$. We are going to fix the free parameter $P_2$ momentarily. We organize the subsequent discussion according to the value of the undeformed mass parameter $M \geq -\frac{1}{8}$. 

\paragraph{$\boldsymbol{8 M = -1}$: Thermal ALD$_{\boldsymbol 3}$.} Making use of eq.~\eqref{eq:kappa3-value} we obtain \begin{equation} 
\label{eq:ald-no-bh-b-on-shell}
    S_{\text{on-shell}} = \frac{\beta_{\text{th}} R_\phi \left(2 P_2 \, n_5 \, R_\phi v_4+n_1 \sqrt{R_\phi^2- n_5 \ap}\right)}{\ap R_{\text{AdS}}} \ . 
\end{equation}
We can fix the parameter $P_2$ by requiring that the on-shell action \eqref{eq:ald-no-bh-b-on-shell} reduce to the AdS$_3$ on-shell action \eqref{S-on-shell-Th-AdS3} in the $R_\phi \to \infty$ limit. We find
\begin{equation} 
\label{eq:ald-b}
    P_2 = -\frac{n_1}{2 n_5 v_4} \ ,
\end{equation}
and \eqref{eq:ald-no-bh-b-on-shell} simplifies to
\begin{equation} 
\label{eq:ald-on-shell}
    S_{\text{on-shell}} = \frac{\beta_{\text{th}} \, R_\phi^2 \, n_1}{\ap R_{\text{AdS}}}  \left(-1+\sqrt{1-\frac{n_5 \ap}{R_\phi^2}}\right) =  \frac{2\pi n_1 R_\phi^2 \tau_2}{\ap} \left(-1+\sqrt{1-\frac{n_5 \ap}{R_\phi^2}}\right) \ , 
\end{equation}
where the second equality is expressed in terms of the modular parameter:
\begin{equation}
    \tau_2 = \frac{\beta_{\text{th}}}{2\pi R_{\text{AdS}}} \ . 
\end{equation}
This gives us a spacetime energy
\begin{equation} \label{eq:e-th-ald}
    E = \frac{\partial S_{\text{on-shell}}}{\partial \beta_{\text{th}}} = \frac{R_\phi^2 \, n_1}{\ap R_{\text{AdS}}} \left(-1+\sqrt{1-\frac{n_5 \ap}{R_\phi^2}}\right) \ .
\end{equation}

\paragraph{$\boldsymbol{-1 < 8 M < 0}$: ALD$_{\boldsymbol 3}$ conical singularity.} This is very similar to the previous case, except that the periodicity of the angle $\phi$ is now $\frac{2\pi}{n}$ for $n\in\mathbb{Z}$. The on-shell action reads
\begin{equation} \label{eq:ald-on-shell-conical}
    S_{\text{on-shell}} = \frac{R^2_\phi \, n_1}{\ap R_{\text{AdS}}} |n| \beta_{\text{cd}} \left(-1+\sqrt{1-\frac{n_5 \ap}{R_\phi^2 n^2}}\right) \ .
\end{equation}
This can again be rewritten in terms of the modular parameter,
\begin{equation}
    \tau_2 = \frac{|n| \beta_{\text{cd}}}{2\pi R_{\text{AdS}}} \ ,
\end{equation}
as
\begin{equation} \label{eq:ald-on-shell-conical-tau2}
    S_{\text{on-shell}} = \frac{2\pi n_1 R_\phi^2 \tau_2}{\ap} \left(-1+\sqrt{1-\frac{n_5 \ap}{R_\phi^2 n^2}}\right) \ .
\end{equation}

\paragraph{$\boldsymbol{M = 0}$: ALD$_{\boldsymbol 3}$ cusp geometry.} Similar to the AdS$_3$ cusp geometry, introducing an additional cut-off at $r=\varepsilon$ for small $\varepsilon>0$ to screen the naked singularity at $r=0$ and taking the $\varepsilon \to 0$ limit gives,
\begin{equation}
    S_{\text{on-shell}} = \frac{\beta_{\text{cusp}} R^2_{\phi} (2 \, P_2 \, n_5\, v_4 +n_1)}{\ap R_{\text{AdS}}} = 0 \ , 
\end{equation}
where in the second equality we used eq.~\eqref{eq:ald-b}. We are including the GHY term at $r=\varepsilon$. This confirms the claim formulated in \cite{Dei:2024sct}.

\paragraph{$\boldsymbol{M > 0}$: ALD$_{\boldsymbol 3}$ black hole.} In this case, the geometry is only defined for $r \geq r_+$. Making use of eq.~\eqref{eq:ald-b}, we obtain
\begin{equation}
\label{S-on-shell-ALD-BH}
    S_{\text{on-shell}} = - \frac{n_1 R_{\phi}^2 \beta_{\text{BH}}}{\ap R_{\text{AdS}}} \left( 1 - \sqrt{1-\frac{4 n_5^2 \pi^2 \ap^2}{R_{\phi}^2 \beta^2_{\text{BH}}}} \right) \ . 
\end{equation}
One can check that eq.~\eqref{S-on-shell-ALD-BH} reduces to eq.~\eqref{S-on-shell-BTZ} in the $R_\phi \to \infty$ limit. Replacing $\beta_{\text{BH}}$ in \eqref{S-on-shell-ALD-BH} by the  expression \eqref{deformed-BTZ-beta}, one can also check that the radicand is positive for any $M>0$. Similarly to the BTZ discussion in Section~\ref{sec:ads3-on-shell-calc}, we can compare our result against expectations from black hole thermodynamics. Using eq.~\eqref{deformed-BTZ-beta}, we find the energy
\begin{equation} 
\label{eq:e-bh-ald}
    E = \frac{\partial S_{\text{on-shell}}}{\partial \beta_{\text{BH}}} = \frac{n_1 R_\phi^2}{\ap R_{\text{AdS}}} \left(-1 + \sqrt{1+\frac{8 M n_5 \ap}{R^2_{\phi}}}\right) \ .
\end{equation}
For large $R_\phi$,  eq.~\eqref{eq:e-bh-ald} correctly reproduces the BTZ black hole mass $\frac{M}{G_3}$. The entropy can be computed in terms of the on-shell action as
\begin{equation}
    \mathcal S = -S_{\text{on-shell}} + \beta_{\text{BH}} \partial_{\beta_{\text{BH}}} S_{\text{on-shell}} = \sqrt{32 M} \pi n_1 n_5 \ ,  
\end{equation}
where in the second equality we used \eqref{deformed-BTZ-beta}. This agrees with the Bekenstein-Hawking expectation,
\begin{equation}
\label{A/4G-ald}
    \mathcal S = \frac{2\pi A}{ \kappa_3^2 e^{2 \Phi(r_+)} } \ ,
\end{equation}
where 
\begin{equation}
    \frac{1 }{8 \pi}  \kappa_3^2e^{2 \Phi(r_+)}
\end{equation}
is the local notion of gravitational constant and 
\begin{equation}
    A = \pi \sqrt{ \frac{32 M n_5 \ap}{1+\frac{8 M n_5 \ap}{R_\phi^2}} } \ ,
\end{equation}
is the horizon area computed in the string frame. Notice that eq.~\eqref{A/4G-ald} correctly reproduces eq.~\eqref{eq:entropy-ads3}.

\paragraph{Spacetime mass.}
In each above case, the spacetime mass reads
\begin{equation}
\label{ALD-mass}
    E = \frac{n_1 R_\phi^2}{\ap R_{\text{AdS}}} \left(-1 + \sqrt{1+\frac{8 M n_5 \ap}{R^2_{\phi}}}\right) \ , 
\end{equation}
which agrees with the result of \cite{Chang:2023kkq}; see their eqs.~(4.27) and (4.28).

\subsection[Holographic interpretation and $\TT$]{Holographic interpretation and $\boldsymbol{T \overline T}$} 
\label{sec:comparison}

In this section, we discuss the holographic interpretation of the on-shell action that we found in the previous section. As discussed in Section~\ref{sec:introduction}, we expect the on-shell action to agree with a CFT$_2$ energy deformed by an irrelevant operator with properties similar to a $\TT$ deformation. If we take the usual double trace $\TT$ deformation then the deformed energy reads \cite{Zamolodchikov:2004ce,Smirnov:2016lqw,Cavaglia:2016oda} 
\begin{equation} \label{eq:e-lambda-ttbar}
    E = \frac{L}{2 \lambda} \left(-1 + \sqrt{1+ \frac{4 \lambda E_0}{L}} \right) \ ,
\end{equation}
He:2025ppzwhere $E_0$ is the undeformed energy, $\lambda$ is the deformation parameter, and $L$ is the radius of the cylinder; for reviews of $\TT$, see~\cite{Jiang:2019epa, He:2025ppz}. From the AdS$_3$ limit, we see that $L = R_{\text{AdS}}$. Comparing with \eqref{ALD-mass}, we deduce
\begin{equation} \label{eq:ttbar-lambda}
	\lambda = \frac{\ap R_{\text{AdS}}^2}{2 n_1 R_\phi^2} = \frac{\ap^2 n_5}{2 n_1 R_\phi^2} \ ,
\end{equation}
and also
\begin{equation}
    E_0 = \frac{4 M n_1 n_5}{R_{\text{AdS}}} \ .
\end{equation}
As a cross-check: taking the limit to thermal $\AdS_3$ we see that
\begin{equation}
    E_0 = -\frac{c}{12 R_{\text{AdS}}} \ ,
\end{equation}
where $c$ is the Brown-Henneaux central charge given in \eqref{eq:brown-c}.

However, the identification of $\lambda$ in eq.~\eqref{eq:ttbar-lambda} assumes a double trace $\TT$ deformation. If we extrapolate our answer to the case $n_5=1$, then we have a precise correspondence where the deformation is actually a single trace $\TT$ deformation of a symmetric product CFT \cite{Dei:2024sct}. This is not true for $n_5>1$. If we assume that the energies of the deformed theory scale in a way suggested by the single trace structure, then we would rescale the deformation parameter  by a factor of $n_1$ to
\begin{align} \label{singletrace}
    \lambda = \frac{\ap^2 n_5}{2 R_\phi^2} \ .
\end{align}
What this highlights is that the precise identification of the field theory deformation parameter requires a better understanding of how the CFT$_2$ is deformed for $n_5>1$ with the constraint that the identification reduces to \C{singletrace} for the $n_5=1$ case.

\section*{Acknowledgements} 
We would like to thank 
Soumangsu Chakraborty, 
Chih-Kai Chang, 
Chris Ferko,
Matthias Gaberdiel,
Bob Knighton,
Ji Hoon Lee,
Emil Martinec 
and especially 
Bob McNees 
for useful discussions and correspondences. We also thank  
Chris Ferko, Emil Martinec 
and 
Bob McNees 
for their helpful comments on a draft of this paper. 

A.~D. is supported by a Reinhard and Mafalda Oehme Postdoctoral Fellowship in the Enrico Fermi Institute at the University of Chicago. K.~N. is supported by a grant from the Swiss National Science Foundation (SNSF) as well as the NCCR SwissMAP. S.~S. is supported in part by NSF Grant No. PHY2014195 and NSF Grant No. PHY2412985.

\bibliography{bib.bib}
\bibliographystyle{JHEP.bst}

\end{document}